# Title: A Knotted Meta-molecule with 2-D Isotropic Optical Activity Rotating the Incident Polarization by 90°

**Authors:** Wending Mai[1,2], Lei Kang[3], Chunxu Mao[3], Ronald Jenkins[3], Danny Zhu[4], Pingjuan Werner[3], Douglas H. Werner[3], Jun Hu[5], Weiping Cao[2], and Yifan Chen[1]

**Affiliations:**

[1] School of Life Science and Technology, The University of Electronic Science and Technology of China, Chengdu, 611731 China

[2] School of Information and Communication, Guilin University of Electronic Science and Technology, Guilin, 541004 China

[3] Electrical Engineering Department, The Pennsylvania State University, University Park, PA 16802 USA

[4] United States Military Academy, West Point, NY 10996 USA

[5] School of Electronic Science and Engineering, The University of Electronic Science and Technology of China, Chengdu, 611731 China

Correspondence to: Wending Mai wdm@ieee.org and Yifan Chen yifan.chen@uestc.edu.cn

**Main text:**

Optical activity is the ability of chiral materials to rotate linearly-polarized (LP) electromagnetic waves. Because of their intrinsic asymmetry, traditional chiral molecules usually lack isotropic performance[1–4], or at best only possess a weak form of chirality[5–9]. Here we introduce a knotted

[14-27] chiral meta-molecule that exhibits optical activity corresponding to a 90° polarization rotation of the incident waves. More importantly, arising from the continuous multi-fold rotational symmetry of the chiral torus knot structure, the observed polarization rotation behavior is found to be independent of how the incident wave is polarized. In other words, the proposed chiral knot structure possesses two-dimensional (2-D) isotropic optical activity as illustrated in Fig. 1, which has been experimentally validated in the microwave spectrum. The proposed chiral torus knot represents the most optically active meta-molecule reported to date that is intrinsically isotropic to the incident polarization.

In 1811, scientists discovered that some substances were able to rotate the polarization of linearly-polarized light. This phenomenon is known as optical activity, or polarization rotation[1–4,9,10], which has been widely applied to the measurement of chiral molecule concentrations in compounds and solutions. These materials, from a macroscopic perspective, usually exhibit isotropic optical activity due to the random orientation of their constituent chiral particles. Nevertheless, most known strongly chiral molecules lack symmetry, thus intrinsically they do not exhibit isotropic optical activity. As symmetry is of fundamental importance in physical chemistry for its relationship with dipole moment, spectroscopic transitions and molecular orbitals, molecules that possess both symmetry and strong chirality are potentially transformative but have thus far been elusive. Based on resonant architectures that are non-superimposable on their own mirror images, researchers have realized a variety of 2-D optically active metasurfaces composed of meta-molecules that are both rotationally symmetric and chiral. However, such thin chiral patterns show rather weak chirality, including a correspondingly small polarization rotation angle[5–9]. Niemi *et al.* proved analytically that, in order to develop a twisted polarizer that rotates the polarization by 90°, 3-D components with strong chirality are required[11]. The helix, a widely used 3-D chiral structure for polarization transformation, has been employed for the realization of twist polarizers. However, the helix, as well as many other known 3-D chiral structures, lacks symmetry[11]. In order to improve symmetry, it was further suggested that several separate 3-D chiral structures should be arranged in a rotationally symmetric fashion. However, meta-molecules based on disjointed structures (*e.g.*, pairs of helices with unconnected ends) generally exhibit a certain chirality transition due to their geometrical discontinuities, which can result in variable optical activity that is highly dependent on the polarization angle of the incident waves[12,13]. Therefore, although highly efficient polarization rotation was observed for a pair of orthogonal polarizations, these well-arranged helix based metamaterials still exhibit incident polarization dependent optical activity[8,12,13]. A meta-molecule that is intrinsically isotropic and simultaneously exhibits strong optical activity has not yet been reported.

Knots represent a unique class of continuous 3-D structures that are capable of exhibiting strong chirality. A knot is a closed curve in three-dimensional space that does not intersect itself anywhere [14,15]. Its applications have been enriching our understanding of a variety of physical phenomena in fluid dynamics [16], photonics [17], and molecular biology (*e.g.*, DNA) [18–20]. Here a parameterized torus knot structure is studied which possesses both chirality and rotational symmetry. A torus knot lies on the surface of a circular or elliptical torus in $\mathbf{R}^3$. The parametric equations that describe the family of circular torus knots are given by:

$$\begin{cases} x = (a + b \cdot \cos(q \cdot t)) \cdot \cos(p \cdot t) \\ y = (a + b \cdot \cos(q \cdot t)) \cdot \sin(p \cdot t) \\ z = b \cdot \sin(q \cdot t) \end{cases} \quad (1)$$

where *a* represents the distance from the origin to the center of the torus tube, while *b* is the radius of the torus tube, and $0 \leq t \leq 2\pi$. The integers *p* and *q* are co-prime (*i.e.*, their greatest common divisor is 1), which indicate the number of times the knot wraps around the torus in the longitudinal and meridional directions, respectively. A longitude curve runs the long way around the torus, while a meridian curve runs the short way. Although a knot and its deformations are considered as topologically equivalent in knot theory, in the field of electromagnetics, deformations will dramatically change the surface current distribution, rendering quite different scattering responses. The electromagnetic scattering properties of perfectly conducting thin knotted wires have been studied in [21-27].

All non-trivial (*p, q*)-torus knots ($p \neq 1, q \neq 1$) possess both chirality and *q*-fold rotational symmetry ($C_q$), which suggests the possibility of intrinsic isotropic optical activity. It is expected that a larger value of *q* will yield a higher degree of symmetry. Here we consider the case where $q = 5$, which is large enough to demonstrate isotropic optical activity while still being relatively easy to fabricate. We choose the parameter $p = 2$, which is the minimum requirement for a non-trivial knot. As shown in Fig. 2 (A), the knot can be viewed as comprising a continuous combination of five segments of a helical structure, which gives rise to its inherently strong chirality and associated ability to rotate the polarization by 90°. Moreover, instead of aligning several separate structures in a rotational symmetric fashion, the entire knot possesses a unified $C_5$ symmetry, resulting in its 2-D isotropic optical activity.

In contrast to a conventional helix, the chirality of a knot depends only on the over and under crossings, not the orientation of the curve. Using the procedure described in [28] together with the expressions defined in Eq. (1), it follows that the knotted meta-molecule shown in Fig. 2(A) has five '-' crossing nodes, with left-handedness. The right-handed mirror image of this structure is shown in Fig. 2 (B), which has 5 '+' nodes. For simplicity and without loss of generality, only the left-handed knotted chiral meta-molecule is studied in this paper.

For validation, optical activity is classically measured by a polarimeter, which contains a polarizer that allows waves with only a specific polarization angle to pass through, a tube that contains the specimen under test, and an analyzer to detect the polarization rotation angle. In order to simply the fabrication and characterization process, a proof-of-concept knot design was developed for operation in the microwave spectrum. For this purpose, the parameters of the torus knot defined in Eq. (1) were chosen to be: $a = 8$ mm, $b = 4$ mm, $c = 4$ mm, and the diameter of the knotted wire was 1.6 mm. Note that the diameter of the knotted wire will affect the gap distance in the central region, which impacts the coupling and the corresponding degree of polarization rotation. Therefore, it is essential that this parameter be carefully optimized to achieve the best performance. The proposed knotted meta-molecule is placed in the center section of a circular waveguide, which acts as the polarimeter tube (see Fig. 3). The torus knot is fixed upon a plastic supporting tray in the circular waveguide, which allows it to be illuminated by normally incident electromagnetic waves. Both ends of the circular waveguide are connected to adaptors that transform the circular waveguide into a rectangular waveguide. The $TE_{10}$ mode dominates in the rectangular waveguide, with the electric field oriented in one specific polarization direction. Correspondingly, the dominate mode in the circular waveguide is the $TE_{11}$ mode, with the polarization mainly oriented in the same direction as that in the adjacent rectangular waveguide. As a result, these two adaptors can function as both the polarizer and the analyzer for the microwave polarimeter. Here we employ an approximation that, at the center of the circular waveguide, the polarization of the $TE_{11}$ mode is mainly oriented in one direction, similar to a LP wave.

The proposed knotted meta-molecule was fabricated by employing a Selective Laser Melting (SLM) technique, using 925 silver (Fig. 4 (A)). The microwave polarimeter (*S1*) was fabricated with aluminum and subsequently electroplated by silver. Both ends of the polarimeter were connected to a Vector Network Analyzer (VNA) through SMA adaptors and cables (Fig. 4 (B)). By extracting *S*-parameters from the two wave-ports of both adaptors, it is possible to detect energy transmission and reflection. Then, by rotating the waveguide adaptors on the two ends relative to one another, the degree of polarization rotation can be extracted by measuring the amount of energy transferred.

As shown in Fig. 4 (B), the polarimeter operates with its two ends perpendicular to each other. Through the transmission of the cross-polarized wave component, it was possible to investigate the ability of the meta-molecule to rotate the incident polarization by 90°. Measured results are shown in Fig. 5. In the absence of the meta-molecule, the transmission of cross-polarized waves is almost zero, since no polarization rotation was introduced. However, when the meta-molecule is inserted, the transmission of the cross-polarized field component at 4.815 GHz is nearly 0.9. This result indicates that the proposed meta-molecule rotates the incident polarization by 90º with an extremely high efficiency.

Next we rotate the polarimeter tube, as well as the knotted meta-molecule placed inside, in order to demonstrate the 2-D isotropic optical activity of the knot. Fig. 5 (A) and (B) show that at a specific frequency (4.815 GHz), the cross-polarization transmission and reflection performance are insensitive to the rotation angle of the knot. In Fig. 5 (C), we show in detail the cross-polarized transmission and reflection at 4.815 GHz with different rotation angles of the knot. In particular, at 4.815 GHz the cross-polarized transmission is over 0.8, with a small reflection less than 0.3, which is insensitive to the rotation of the knotted meta-molecule. This result experimentally confirms the transformative property that the polarization rotating ability of the proposed knotted meta-molecule is independent of the incident polarization angle. As mentioned previously, one could expect even better performance for a knotted meta-molecule with a larger value of the parameter $q$.

The proposed meta-molecule exhibits 2-D isotropic optical activity due to its intrinsically strong chirality and the continuous multi-fold rotational symmetry of the torus knot. The exotic molecular symmetry of the torus knots calls for further studies on their potential for a wide range of applications in optical physics, chemistry, material science, and engineering.

S1 is included in the Supplementary Material.

**Acknowledgments:** This work was supported in part by the EPSRC Program (EP/R013918/1), the National Excellent Youth Fund by NSFC (61425010), the Foundation for Innovative Research Groups of NSFC (61721001), the Changjiang Scholar Program, the Penn State MRSEC, Center for Nanoscale Science (NSF DMR-1420620), and the Sichuan Science and Technology Program (2018GZ0251).

The opinions in this work are solely of the authors and do not necessarily reflect those of the U.S. Military Academy, the U.S. Army, or the Department of Defense.

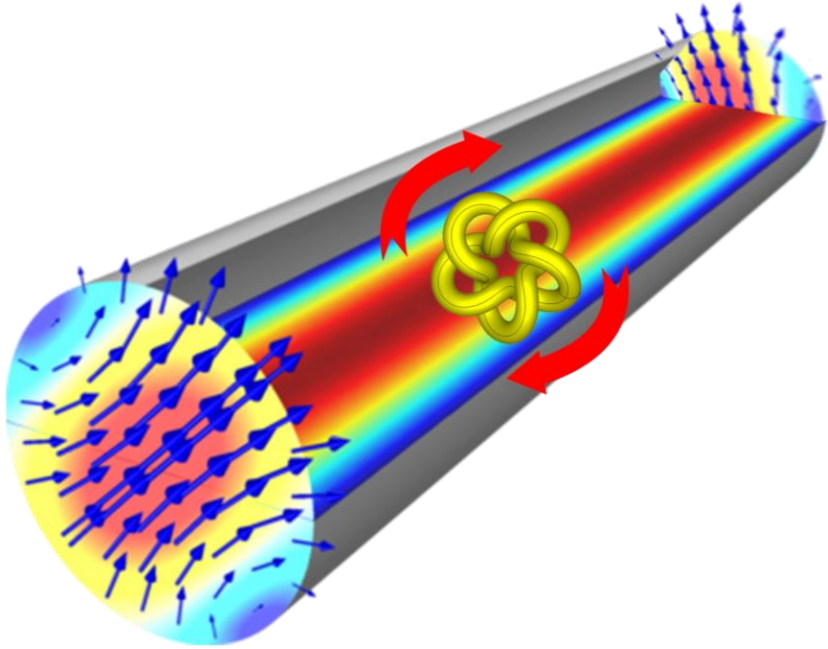

**Fig. 1.** The proposed knotted meta-molecule can rotate the incident linear polarization by 90° with high efficiency. Moreover, because of its continuous multi-fold rotational symmetry, the optical activity of the knotted meta-molecule is shown to be independent of the incident polarization angle (*i.e.*, it exhibits two-dimensional isotropy). Namely, rotating the meta-molecule will not affect its optically active performance.

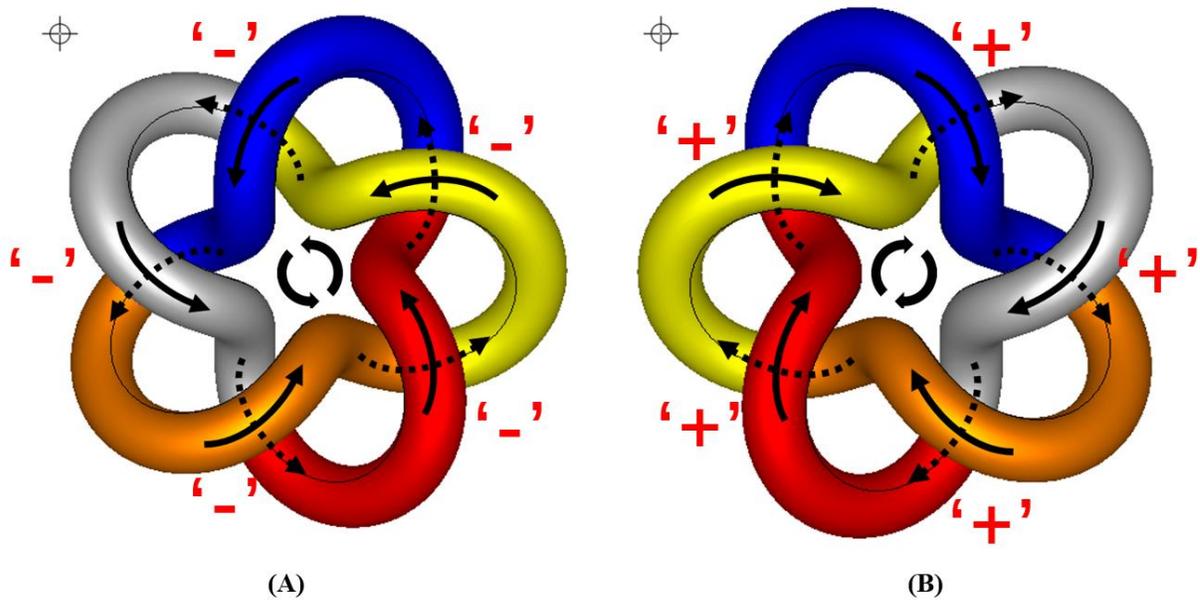

**Fig. 2.** A (2, 5)-torus knot can be viewed as a combination of five helical segments, colored in red, blue, orange, yellow and silver, respectively. Given a linear polarized wave normally incident from the front, the **(A)** left-handed and the **(B)** right-handed knotted meta-molecule can rotate the polarization counter-clockwise (levorotary) and clockwise (dextrorotary), respectively.

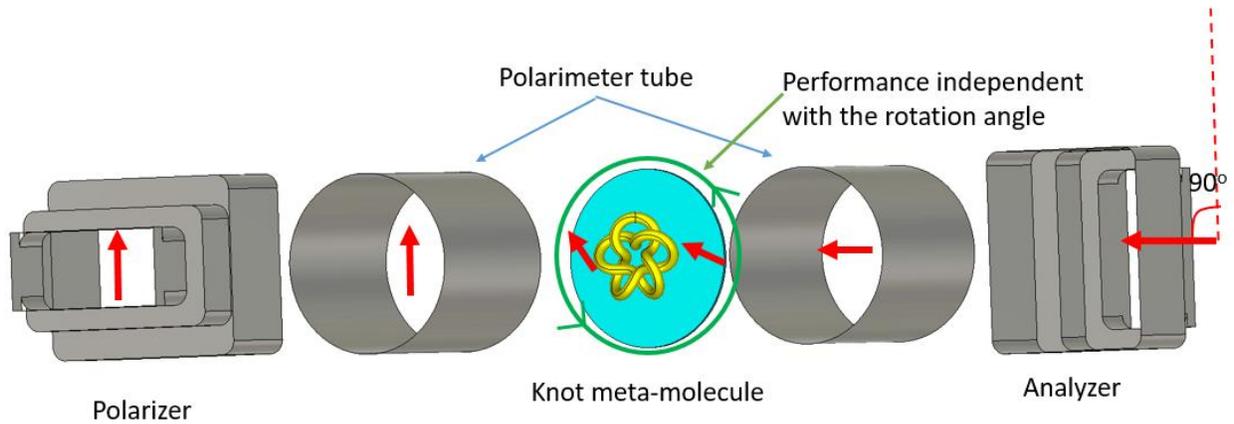

**Fig. 3.** The polarization rotation is measured by a microwave polarimeter which contains a polarizer ($TE_{10}$ mode) allowing waves with only a specific polarization angle to pass through, a tube ($TE_{11}$ mode) with the specimen under test, and an analyzer ($TE_{10}$ mode) to detect the polarization rotation angle.

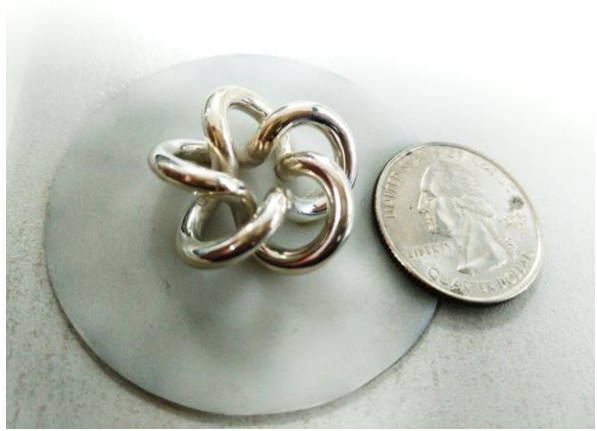

**(A)**

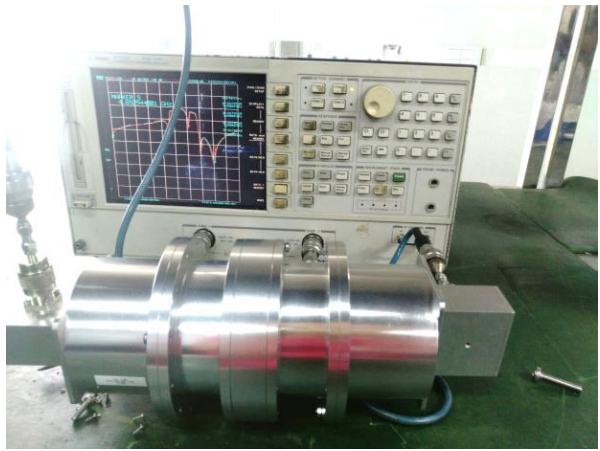

**(B)**

**Fig. 4.** The fabrication and measurement setup of the knotted meta-molecule. **(A)** The knotted meta-molecule is fabricated using silver with a metal SLM technique. **(B)** The microwave polarimeter is connected to a VNA in order to extract transmission and reflection results.

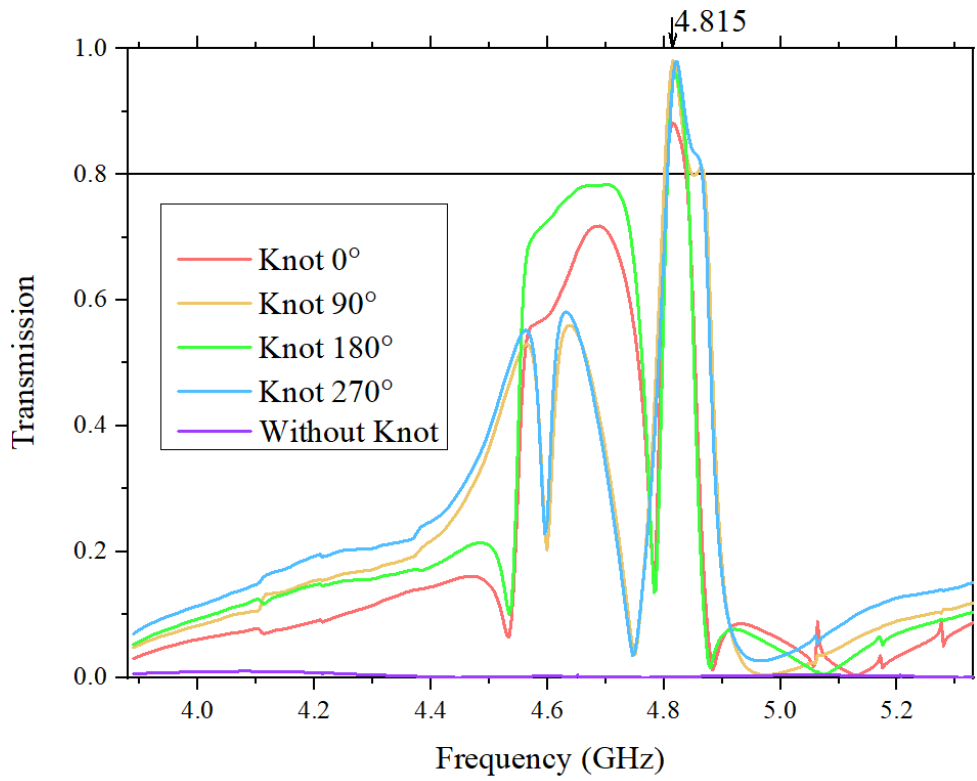

**(A)**

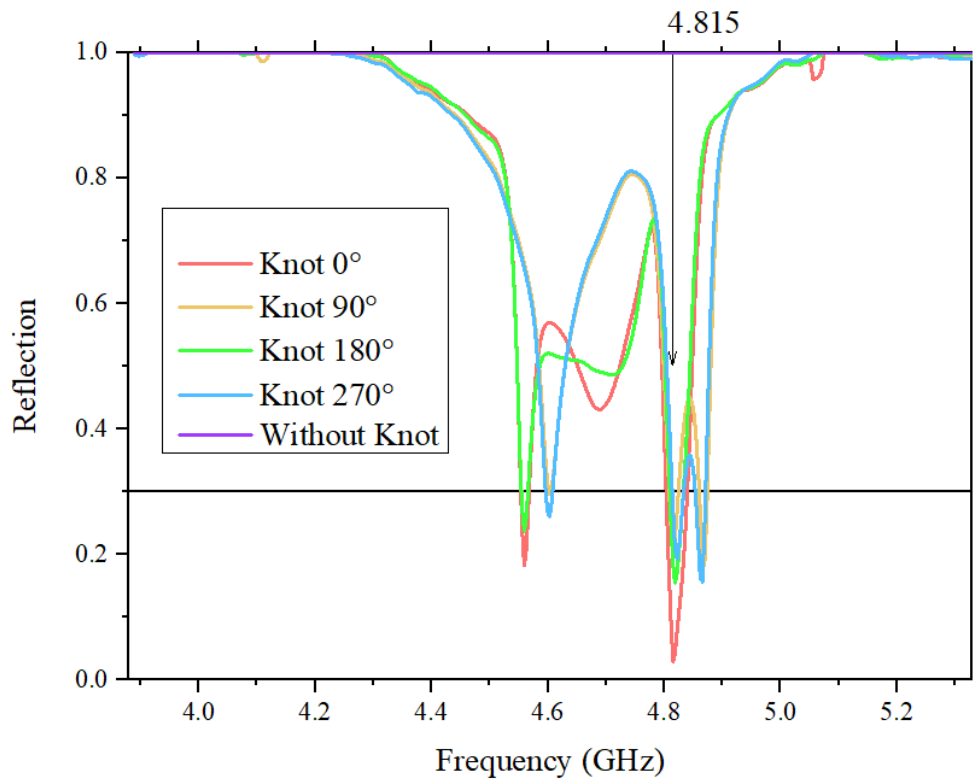

**(B)**

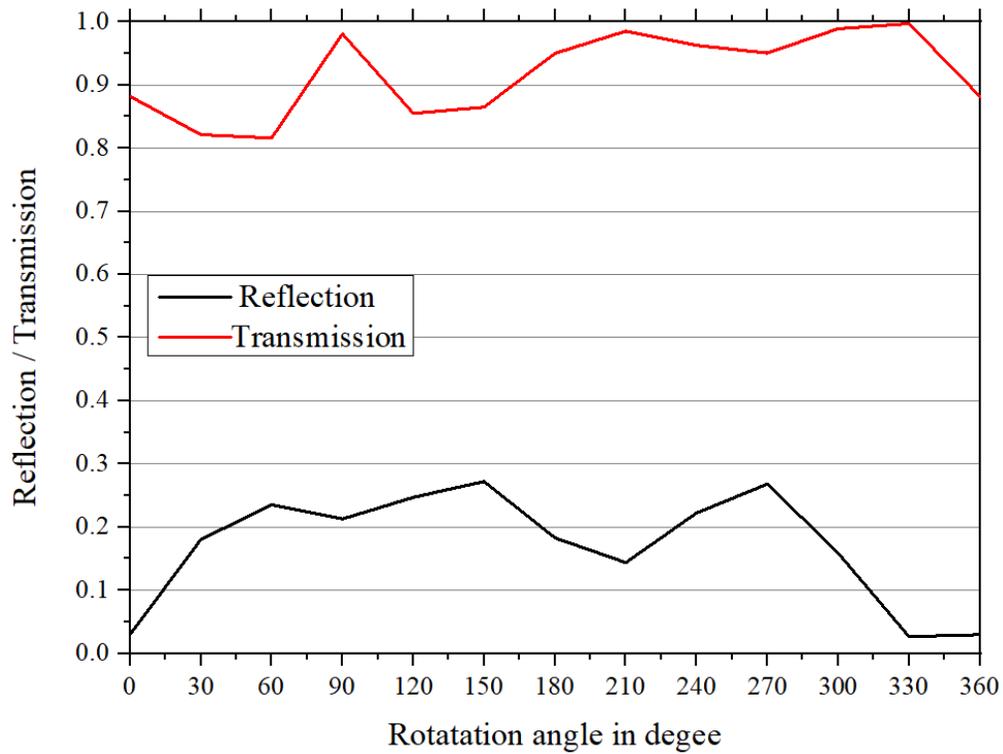

**(C)**

**Fig. 5**. The polarization rotation measurement of the knotted meta-molecule. When rotating the knotted meta-molecule, the transmission **(A)** and reflection **(B)** of the cross-polarization at 4.815 GHz are insensitive to rotation of the knot. **(C)** Transmission and reflection at 4.815 GHz with rotation angle scan at every 30°.

# Supplementary Materials:

## S1.

## Design for the Microwave Polarimeter

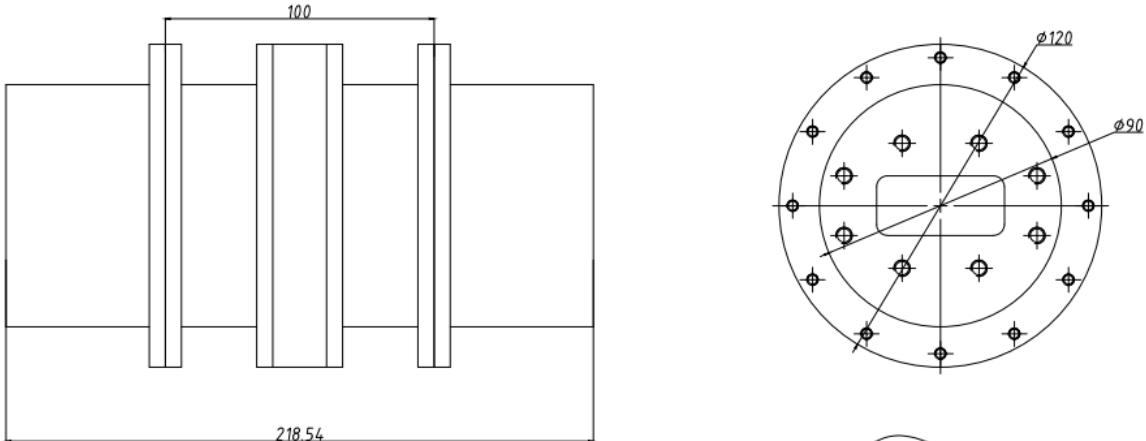

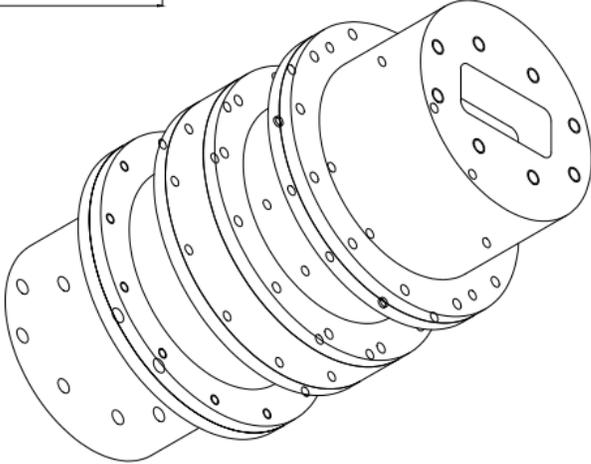

(A)

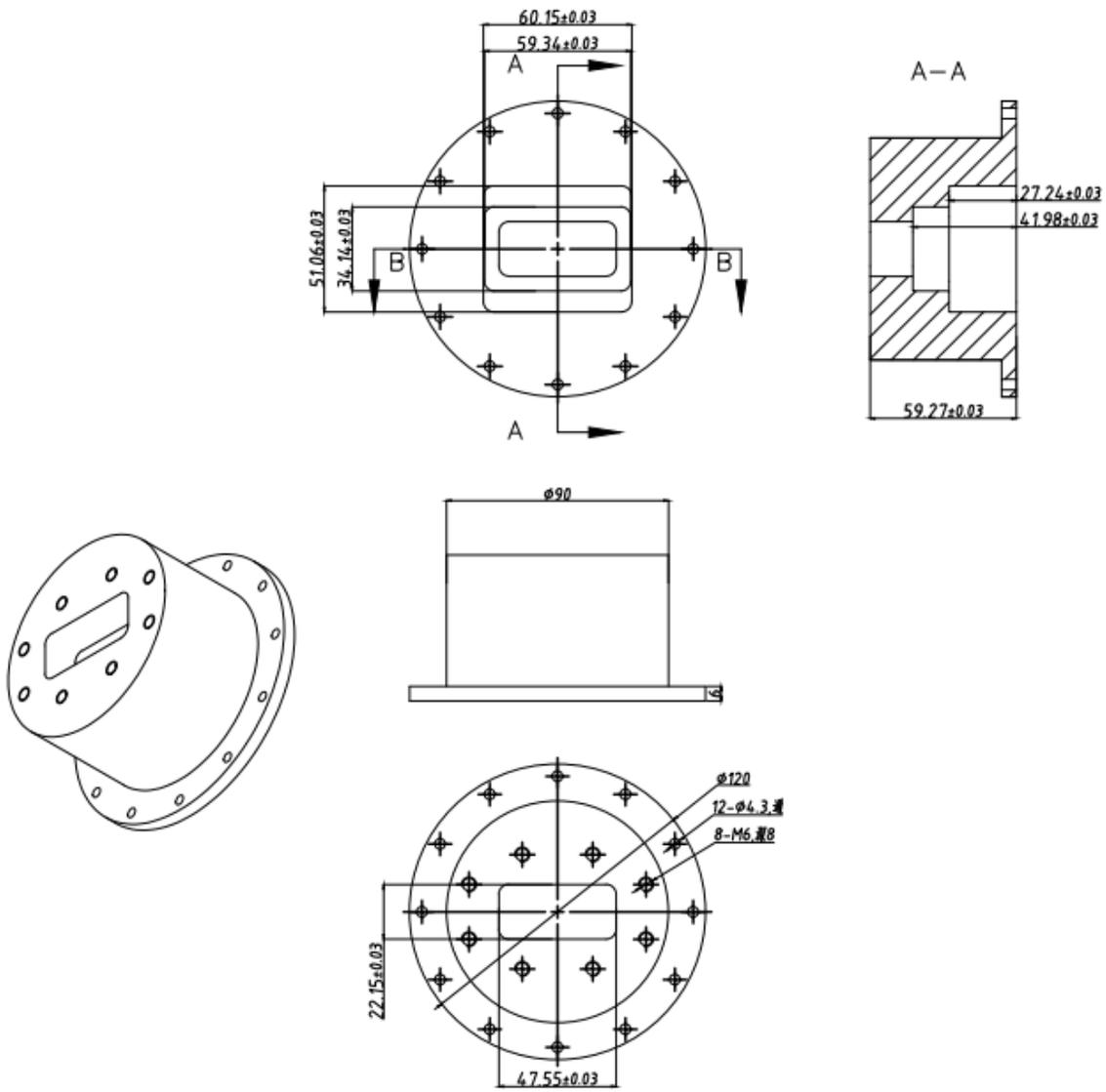

(**B**)

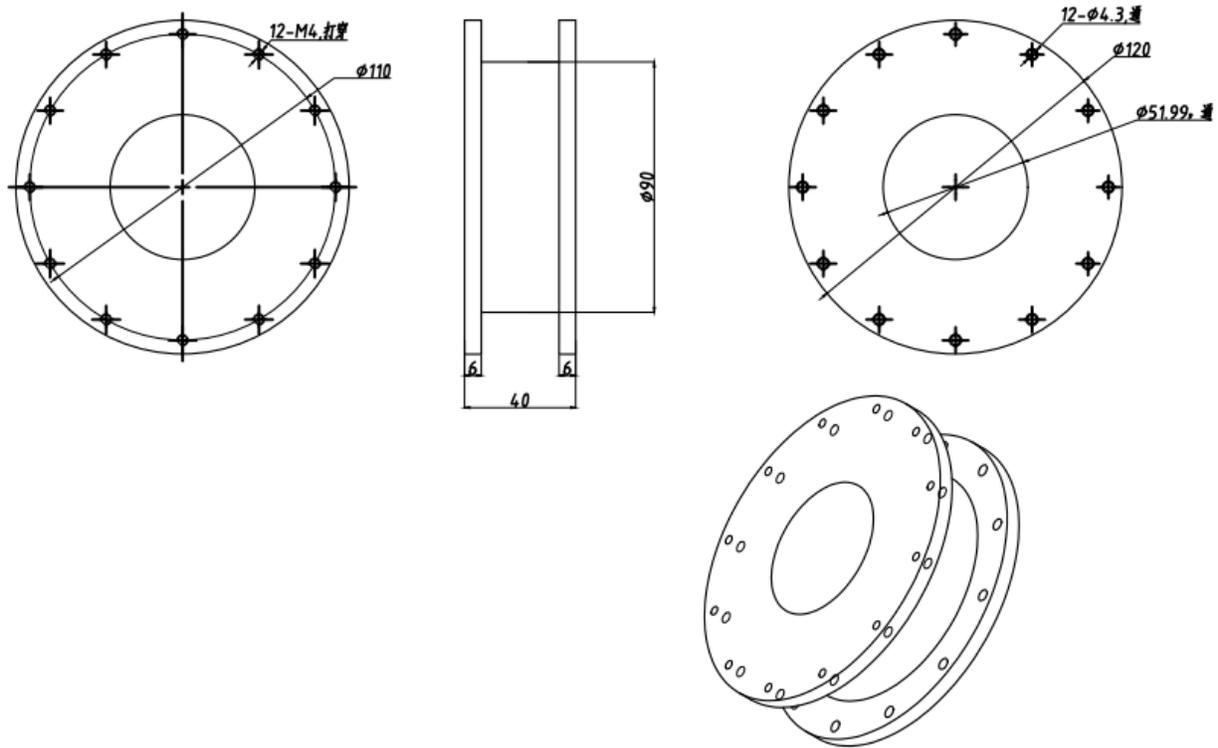

(C)

**Fig. S1.** The design details for the microwave polarimeter. **(A)** Outline of the assembled polarimeter. **(B)** The rectangular-circular waveguide adaptor, which functions as the polarizer and the analyzer. **(C)** A segment of circular waveguide functionalized as the polarimeter tube. There are many holes that are positioned to allow the rotation of the tube and the specimen under test in increments of 30°.